# Computational Study of Defect variant Perovskites $A_2BX_6$ for Photovoltaic Applications

M. Faizan[1,2*], K. C. Bhamu[3,4], S. H. Khan[1], G. Murtaza[5], Xin He[2]

[1]Department of Physics, University of Peshawar, Peshawar 25120, Pakistan

[2]State Key Laboratory of Superhard Materials and School of Materials Science and Engineering, Jilin University, Changchun 130012, China

[3]Physical and Materials Chemistry Division, CSIR-National Chemical Laboratory, Pune 411008, India

[4]Department of Physics, Gramin Mahila P.G. College, Sikar 332024, Rajasthan, India

[5]Materials Modeling Lab, Department of Physics, Islamia College University, Peshawar 25120, Pakistan

*Correspondence: faizanstd@uop.edu.pk

**ABSTRACT:** A comprehensive study of the structural, electronic, and optical properties of lead-free perovskites has been carried out by means of first principles method based on DFT. The calculations are performed for the compound of the type $A_2BX_6$ with A = Rb, and Cs; B = Sn, Pd, and Pt; and X = Cl, Br, and I. The calculated structural parameters (lattice constants and bond lengths) agree well with the experiments. The computed band gap reveals a semiconducting profile for all these compounds showing a decreasing trend of the band gap energy by changing the halide ions consecutively from Cl to Br and Br to I. However, for variation in the B-site cation, the band gap increases by changing the cation from Pd to Pt via Sn. The most likely compounds, $Rb_2PdBr_6$ and $Cs_2PtI_6$, exhibit a band gap within the optimal range of 0.9–1.6 eV for single-junction photovoltaic applications. The optical properties in terms of the optimal value of the dielectric constant, optical conductivity, and absorption coefficient are also investigated upto the photon energy of 10 eV. Our results indicate that upon changing the halogens ions (Cl by Br and Br by I) the optical properties altered significantly. Maximum dielectric constant and high optical absorption are found for $Rb_2PdBr_6$, $Rb_2PdI_6$, and $Cs_2PtI_6$. The unique optoelectronic properties such as ideal band gap, high dielectric constants, and optimum absorption of $A_2BX_6$ perovskites could be efficiently utilized in designing high performance single and multi-junction perovskite solar cells.

## 1. INTRODUCTION

Initial studies on lead halide perovskite materials as light absorbers were published in 2009[1-3]. The power conversion efficiencies (PCEs) of these materials are around 24.2% according to a recent report[4]. Such a fast improvement is attributed to the unique photovoltaic properties of Pb halide perovskite absorbers, for example, tunable direct band gap, fair electron and hole effective mass, excellent optical absorption, high stability, benign defect tolerance, and long term photogenerated carrier diffusion lengths[5-7]. The commercial use of perovskites containing Pb for photovoltaic applications has attracted a great deal of research interest[8]. However, two major challenges still exist: lead (Pb) is toxic, and the perovskite suffer from low chemical stability in air[9-18]. The toxic nature of Pb causes health problems such as narcosis and irritation of eyes, nose, and throat[19]. Likewise, the degradation/instability issue of conventional perovskite materials, especially when retained in a humid environment is highly problematic; upon exposure to moisture, the organo-Pb perovskite, like methylammonium lead iodide ($CH_3NH_3PbI_3$) changes into lead iodide with a total loss of the nitrogen moiety[20-21]. The other major factors that rapidly degrade a perovskite material consist of high temperature, light, and oxygen[19]. Such factors will effect perovskite materials in a variety of ways including decomposition, oxidation, ion diffusion, hydration, and polymorphic transformation[22]. Hence, both the issues (toxicity of lead and chemical instability) are big hurdles that will hold back the large scale productive applications of perovskite solar cells (PSCs) containing lead[19]. Therefore, developing new halide perovskites free of lead for use in PSCs are highly desirable to reduce the toxic issue of the emerging photovoltaic devices. Similarly, many efforts have been made to improve the stability of halide perovskites by using different approaches, including the use of alternative material instead of lead[23-24], fabrication of 2D perovskites[25-26] and using mixed cations[27].

The basic crystal structure of perovskite compounds is of the type $ABX_3$ (in the 1:1:3 ratio) such that A is a monovalent and B a divalent cation, and X is an anion ($CH_3NH_3PbI_3$, $CsPbI_3$, $CsSnI_3$, $CsPbBr_3$)[27-29]. However, some modifications have been made in the basic structure of perovskite, such as $A_2B^{1+}B^{3+}X_6$ (2:1:1:6)[27, 29], $A_3B_2^{3+}X_9$ (3:2:9)[27], and $A_2B^{4+}X_6$ (2:1:6)[27, 30]. Such modification in the structure will also affect the electronic properties of these perovskite compounds[31]. Taking into account both the crystal structure and chemical composition of the perovskite compounds, novel materials for efficient photovoltaic application in the perovskite family could be discovered with the help of computational approach such as the density functional theory[19].

The $A_2B^{1+}B^{3+}X_6$ structure can be considered as cubic by replacing every pair of adjacent B-cations with one $B^{1+}$ and one $B^{3+}$ cation at the top. Typical examples of this kind of the compounds are $Cs_2CuBiBr_6$[32] and $Cs_2AgInBr_6$[33]. The recently discovered double perovskite (DP) halides, $Cs_2AgBiX_6$ (X = Br or Cl)[34] and $(CH_3NH_3)_2BBiX_6$ (B= K, Cu, Ag, Tl; X=Cl, Br, I)[35-36], absorb visible light with high carrier recombination lifetimes, thus giving diverse opportunities for air-stable and nontoxic-alternatives to $Sn^{2+}$ and $Pb^{2+}$ based $ABX_3$ perovskite[34, 37]. The $A_3B_2X_9$ perovskites can be thought of as cubic if by one in the three 'B' cation in $ABX_3$ structure is eliminated at the top. These halide perovskites consist of $BX_6$ octahedral network and the vacant sites between them are filled by the A-site cation. The B-site cation lies at the +3 oxidation state to meet the charge neutrality conditions. Examples of this kind of perovskites are $Cs_3Sb_2I_9$[38-39], $Cs_3Bi_2I_9$[40], and $Rb_3Sb_2I_9$[39]. The $A_2BX_6$ checkerboard pattern can be considered as a derivation from $ABX_3$ structure if half of the B cations at the $[BX_6]$ cluster center are removed[27]. The charge neutrality condition implies that the B-site cation should be a tetravalent, extending the type of cations for substitution into B-site[41]. Typical examples of this kind of perovskites are $Cs_2SnI_6$, $Cs_2PdBr_6$, and $Cs_2TeI_6$[27, 30, 42]. The $A_2BX_6$ compounds could also be derived from the double perovskite $A_2B^{1+}B^{3+}X_6$ by creating a vacancy at the B-site, $A_2\square BX_6$ ($A_2BX_6$, where $\square$ denotes B vacancy)[30] as shown in Figure 1. Generally, it is referred to as antifluorite crystal ($K_2PtCl_6$) and is described by the $[BX_6]^{-2}$ octahedral cluster bridged by the A-site cations[30]. The $A_2BX_6$ structure shows similar features to $ABX_3$ perovskites and most of them possesses cubic structure. In $A_2BX_6$ structure, every other $[BX_6]^{-2}$ octahedra is removed i.e. half of the B-site cation is unoccupied, the close packing behavior similar to $ABX_3$ is maintained[30].

The $A_2BX_6$ with B = Sn and Te have recently been reported capable of absorption of light in the visible to infrared (IR) region giving new hope for stable materials with a nature friendly operation[30, 43-44]. In this framework, $Cs_2SnI_6$ with cubic crystal structure containing Sn in its +4 oxidation state is suitable as a potential candidate for applications in PSCs[43]. Diffuse-reflectance measurements show an optical band gap of 1.25–1.30 eV in comparison with the thin films band gap of 1.60 eV. The density functional theory results in a direct transition nature (at Γ symmetry line) with the suggested band gap (0.13 to 1.26 eV) significantly different from the experimental value given above, most likely using functionals with different correlation approximations[41, 45].

The compound possesses N-type electrical conductivity, strong absorption power, and moisture stability[43-44, 46]. Due to the tetravalent character of Sn, the compound exhibits higher air stability with respect to $CsSnI_3$[41]. In fact, the $Sn^{4+}$ based perovskites, $A_2SnX_6$, show significantly enhanced stability as compared to $Sn^{2+}$ based $ABX_3$ perovskites. The substitution results in performance enhancement in optoelectronic devices such as the light emitting diodes, the flexible electronic components, and photodetectors[30, 43, 47]. In fact, numerous non- or low-toxic transition metals have stable +4 oxidation state paving the way for finding favorable halide perovskites; for example by replacing the $Sn^{4+}$ in $Cs_2SnI_6$ by appropriate transition-metal cations[48]. This is established by an ongoing report by Sakai et al. who studied $Cs_2PdBr_6$ as a novel perovskite for use in PSCs[42]. The optical band gap of $Cs_2PdBr_6$ calculated from absorption measurement was 1.60 eV[42]. The effective masses of the electron and hole calculated from first principles calculations were reported to be 0.53 $m_e$ and 0.85 $m_e$, which indicate the N-type semiconducting behavior of $Cs_2PdBr_6$[42]. The compound exhibited superior stability and is advantageous for optoelectronic applications[41-42]. Ju et al. carried out an integrated experimental and theoretical study of Ti-based vacancy ordered DPs $A_2TiX_6$ (A = $K^+$, $Rb^+$, $Cs^+$ $In^+$; X = I, Br, or Cl) and $Cs_2TiI_xBr_{6-x}$, showing a suitable band gap in the range from 1.38 eV to 1.78 eV for photovoltaic applications[48]. Zhao and his co-workers extensively reviewed the progress on Pb-free halide perovskites including different vacancy ordered DPs by combined experimental and theoretical efforts[47]. They studied a new family of vacancy order DPs, $Cs_2BX_6$ (B= Pd, Sn, Ti, Te; X= Cl, I), and claimed that the compounds possess diverse electronic structures and optical features for various optoelectronic applications. Another group of researcher carried out a computational study for seven known compounds of the type $A_2MX_6$, A representing K, Rb, and Cs, M representing Sn, Pd, Pt, Te, and X= Iodine. They used hybrid functional (HSE06) based on DFT[8]. The striking aspects of their study were calculations of electronic structure and finding of the energetic stability. They found that the band gap and effective masses of the compound increase as the A-site cation is varied from K to Rb to Cs.

From the above studies, it is clear that many efforts have been extensively made in the recent past to explore the optoelectronic properties of $A_2BX_6$ type compounds. To conduct further research aimed at the use of various metals substitution for photovoltaic and optoelectronic applications, we make use of the density functional theory to explore new variants in $A_2BX_6$ family with possible A, B, and X as A = Rb, Cs; B = Sn, Pd, or Pt; and X = Cl, Br, or I. In Figure 2, we have shown our selected combination. We begin from the structural properties and then examine the electronic structure as well as optical spectra of these compounds. In addition, we also investigated the thermodynamic stability of these compounds by calculating the formation energies. The main goal of the present work is to search for potential alternatives of lead-free perovskites that preserve the same optoelectronic properties as that of the Pb-based perovskites. We also believe that our work will provide a theoretical support to the future studies of defect-variant perovskites for photovoltaic and other optoelectronic applications.

## 2. COMPUTATIONAL METHODS

Present results were obtained with the help of computational code wien2k[49] based on density-functional theory (DFT)[39]. We employed all electron Full Potential Linearized Augmented Plane Wave (FP-LAPW) method with Perdew, Burke, and Ernzerhof Generalized-Gradient Approximation (PBE-GGA) functional[50] to gain the structural properties, while for electronic and optical properties we use for the first time the most accurate scheme of modified Becke Johnson (mBJ) semi-local exchange potential[51]. We have also performed some additional calculations for the band structures with PBE-GGA, Engel and Vosko (EV-GGA)[52], Gritsenko, Leeuwen, Lenthe, and Baerends functional for solid and correlation (GLLB+SC)[53], GLLB+SC-SOC, and mBJ+SOC methods. The mBJ has been shown to produce more accurate band gap as compare to standard LDA/GGA functionals[54-55]. It is an orbital independent potential that yields accurate band gap for a large number of solids including insulators, semiconductors and strongly correlated transitional metal oxide[51, 55-56]. The relaxation of the size, shape and the relative atomic positions of the unit cells was done with an energy cutoff of 400 Ryd for plane wave expansion and ended when the energy become within $10^{-4}$ Ryd. For self-consistent calculations, the convergence criteria for the charge and atomic force were set at 0.001 $e$ and 0.05mRy/a.u, respectively. The convergence with respect to basis size is through the parameter, $R_{MT}K_{max}$, which is the product of the smallest muffin-tin sphere radius ($R_{MT}$) times the largest plane wave vector ($K_{max}$). For calculations of the electronic and optical properties we used 5000 k-points. The other parameters are $G_{max} = 12\ a.u^{-1}$ (plane wave cutoff) and muffin-tin radii for Rb, Cs, Sn, Pd, Pt, Cl, Br, and I atoms are selected as 2.5 a.u for Rb, 2.5 a.u for Cs, 2.07 a.u for Sn, 2.11 a.u for Pd, 2.21 a.u for Pt, and 1.71/1.83/1.91 a.u for Cl/Br/I respectively. For all the compounds, the lattice constants and atomic positions from experimental data were used to perform the calculations. Calculations with finer k-points and high energy cutoffs confirmed the convergence of the results on the lattice parameters and electronic band structure.

## 3. RESULTS AND DISCUSSION

### 3.1. Crystal Structure and Ground State Properties

The vacancy ordered DPs $A_2BX_6$ (A= Rb and Cs; B = Sn, Pd, and Pt; and X= Cl, Br, and I) have face centered cubic structure with space group $Fm\bar{3}m$ (No. 225). The atomic positions and geometric configuration of $A_2BX_6$ is illustrated in Figure 1 which can be described as B-deficient $ABX_3$ perovskites with $[BX_6]$ cluster. The vacant sites between $[BX_6]$ octahedra are filled with A-site atoms. Each $[BX_6]$ octahedra in $A_2BX_6$ structure is isolated from one another forming a 12-fold coordination environment of discrete X anions. As a result, the bond length B-X tend to get shorter in the $A_2BX_6$ perovskites[57]. Figure 1 shows that A-cations are surrounded by twelve halogens ions, B atoms are surrounded by six halogens ions, located in the center of octahedron. Each $BX_6$ octahedra is centered in a cubic environment, placed at the corners and at the face center positions. The A-site atoms are located at 8c Wyckoff site and (1/4, 1/4, 1/4) fractional coordinates, B-site cations are located at 4a Wyckoff site and (0, 0, 0) coordinates, and X-anions are sited at 24e Wyckoff site and (x, 0, 0) fractional coordinates. The variable x lies around 0.2 for each of the compound. The optimization curve achieved by fitting the Murnaghan's equation of state (EOS) to the PBE-GGA's total energies verses volume data for the representative candidate is shown in Figure 3. The relaxed internal parameters obtained by doing total energy minimization give an excellent measure of nearest neighbor distances and the resultant electronic structure. The predicted lattice parameters obtained after geometry optimization along with experimental results are listed in Table 1. The obtained results clearly suggest better performance of PBE-GGA in predicting structural properties of materials. The calculated lattice constants are found in good agreement with experiments showing an increasing trend with changing Cl to Br and then to I (Figure 4). The unit cell volume is found to increase when the halogen atom (X) is changed by Cl to Br to I. In structural relaxation, the $[BX_6]$ octahedra shrink in a regular fashion leading to a smaller B-X bond lengths than the simple perovskites[57]. The bond lengths obtained after energy minimization for $A_2BX_6$ compounds are presented in Table 2 along with the available experimentally measured bond lengths. Table 2 indicates that the bond lengths A-X and B-X show an increasing trend by increasing the size of the halogen ion. Among the eight compounds, the Rb$_2$SnI$_6$ possesses larger bond lengths and thus has a larger lattice parameter.

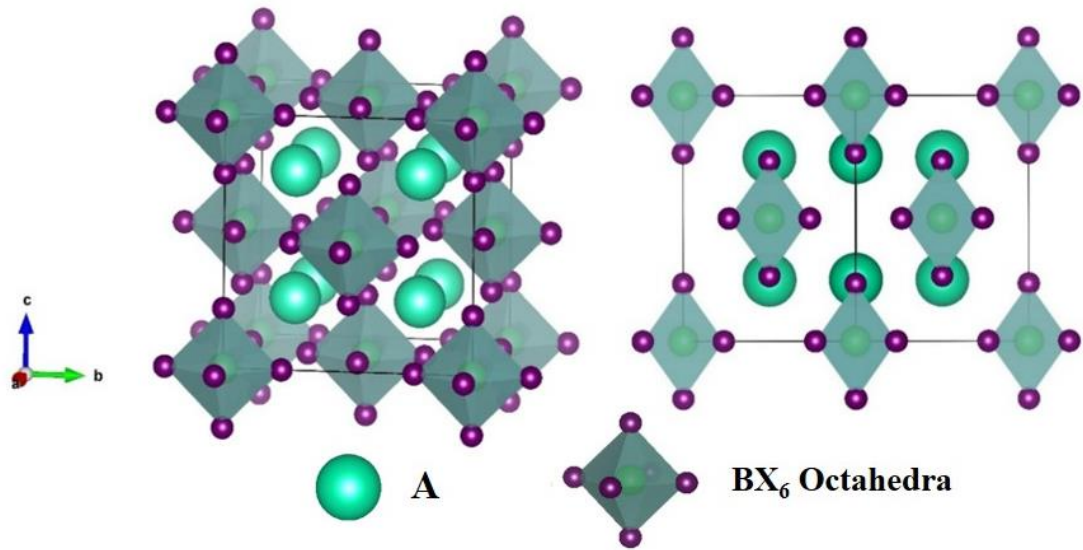

**Figure 1**. Schematic crystal structures of A$_2$BX$_6$ compounds in cubic Fm$\overline{3}$m (Left) and reorientation of the unit cell (right). The BX$_6$ octahedra are shaded, with the X-atom on the corners. The A-cations are in the hole between BX$_6$ octahedra.

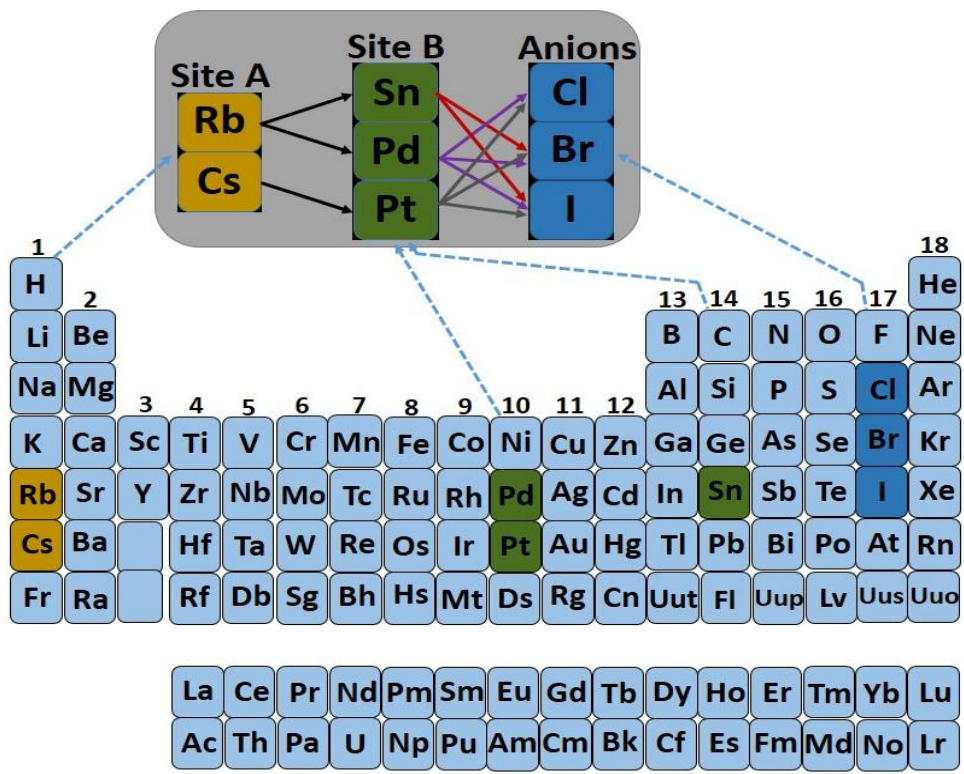

**Figure 2.** Selection of elements for site A, B, and anion X with composition $A_2BX_6$.

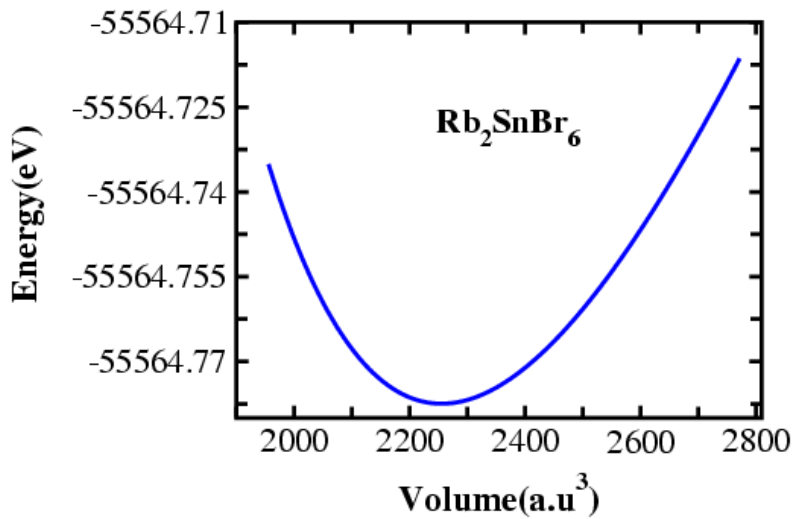

**Figure 3**. The total energy *vs* volume plot for representative $Rb_2SnBr_6$ perovskite calculated with PBE-GGA.

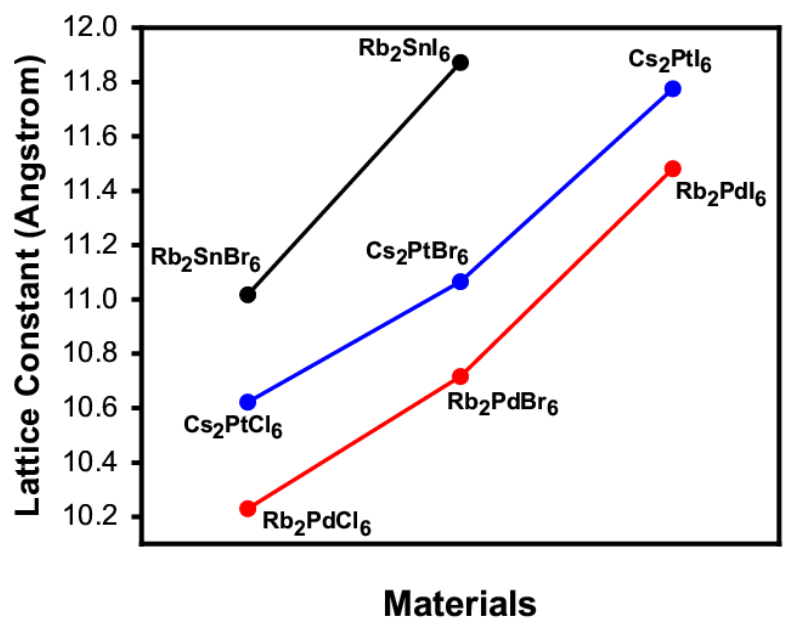

**Figure 4.** Periodic variation of the lattice constants with the halogens atoms.

**Table 1.** Optimized lattice parameters, unit cell volume, and formation energy of $A_2BX_6$ (A= Rb, Cs; B= Sn, Pd, Pt; X= Cl, Br, I) using the PBE functional.

| Compound | a (Å) | Exp. (Å) | Volume/ Å³ | $\Delta H(eV/f.u.)$ |
|---|---|---|---|---|
| $Rb_2SnBr_6$ | 11.01 | 10.58[58] | 1193.98 | -11.08 |
| $Rb_2SnI_6$ | 11.87 | 11.62[59] | 1493.78 | -7.89 |
| $Rb_2PdCl_6$ | 10.23 | 9.990[60] | 956.84 | -10.58 |
| $Rb_2PdBr_6$ | 10.71 | 10.02[61] | 1098.80 | -8.90 |
| $Rb_2PdI_6$ | 11.48 | 11.185[62] | 1351.18 | -6.83 |
| $Cs_2PtCl_6$ | 10.62 | 10.192[63] | 1070.03 | -11.41 |
| $Cs_2PtBr_6$ | 11.06 | 10.670[64] | 1209.78 | -9.63 |
| $Cs_2PtI_6$ | 11.77 | 11.367[65] | 1457.65 | -7.49 |

**Table 2.** Optimized bond lengths of $A_2BX_6$ (A= Rb, Cs; B= Sn, Pd, Pt; X= Cl, Br, I) compounds.

| Compound | A-B(Å) | | A-X (Å) | | B-X (Å) | |
|---|---|---|---|---|---|---|
| | optimized | Exp. | optimized | Exp. | optimized | Exp. |
| $Rb_2SnBr_6$ | 4.7707 | | 3.8966 | | 2.6524 | |
| $Rb_2SnI_6$ | 5.1406 | | 4.1978 | | 2.9061 | 2.85[59] |
| $Rb_2PdCl_6$ | 4.4297 | | 3.6227 | | 2.3513 | |
| $Rb_2PdBr_6$ | 4.6405 | | 3.7928 | | 2.5079 | |
| $Rb_2PdI_6$ | 4.9716 | | 4.0620 | 3.957[62] | 2.7205 | 2.662[62] |
| $Cs_2PtCl_6$ | 4.5996 | | 3.7675 | | 2.3561 | |
| $Cs_2PtBr_6$ | 4.7917 | | 3.9208 | | 2.5096 | |
| $Cs_2PtI_6$ | 5.0989 | | 4.1694 | 4.04[65] | 2.7156 | 2.68[65] |

To assess the thermodynamic stability, we determined the formation energies of $A_2BX_6$ (A= Rb and Cs; B = Sn, Pd, and Pt; and X= Cl, Br, and I) perovskites with respect to the potential pathways. The formation energy of a compound can be defined as the difference between the total energy of the compound and the sum of the energies of the main constituents in its standard form. In the present case, the formation is given by

$$\Delta H_f^{A_2BX_6} = E_t^{A_2BX_6} - 2E_t^{A_{bcc/fcc}} - E_t^{B_{bcc/fcc}} - 6E_t^{X_{ortho}}$$

Where $E_t^{A_2BX_6}$, $E_t^A$, $E_t^B$, and $E_t^X$ are the minimum ground state energies of the $A_2BX_6$, A, B, and X, respectively. For each considered system, we have performed a full geometry optimization i.e. finding the volume for which the material has the lowest energy. In literature, we have obviously noticed that the sign and magnitude of the formation energy is mostly used to characterize the stability of a compounds[66]. The negative formation energy means the material is stable i.e. negative value indicates favorable formation of a compound, whereas a positive value represents instability and material cannot be synthesized at ambient conditions. The larger the magnitude of formation energy higher is the stability of the compound and vice versa. The obtained data of formation energy for the selected candidates is listed in Table 1. Our results demonstrate a negative formation energy showing better stability of these compounds. By comparing the data listed in Table 1, we noticed that $Rb_2SnBr_6$ is the most stable structure among $Rb_2BX_6$ (B = Sn, Pd; X= Cl, Br, I) perovskites. Likewise, the $Cs_2PtCl_6$ perovskite is more stable than $Cs_2PtBr_6$ and $Cs_2PtI_6$ based on their calculated formation energies.

### 3.2. Band Structure and Density of States

Generally, the band gap of inorganic Pb-free perovskites ought to be within the optimal range of 0.9 eV to 1.6 eV, corresponding to an efficiency of solar cell >25%[67]. We investigated the electronic nature of the eight perovskite variants compounds of the type $A_2BX_6$ to see whether they meet this requirement. Band gap calculated with various exchange-correlation functionals of all the compounds are presented in Table 3. The mBJ calculated band gap range from 0.47 eV to 2.91 eV with the lowest band gap of 0.47 eV for $Rb_2PdI_6$ and the highest one of 2.91 eV for $Cs_2PtCl_6$. The calculated mBJ band structures for $A_2BX_6$ are shown in Figures (5-7). Each figure gives the band structure for fixed A-site cation along the highly symmetry direction of the Brillouin zone. A first glance at the computed band structures suggest semiconducting characteristics of all these compounds. Starting with the Sn-based compounds, $Rb_2SnBr_6$ and $Rb_2SnI_6$, both the systems are predicted to be semiconductor with direct band gap nature (Figure 5). The valence band maximum (VBM) and conduction band minimum (CBM) are localized at the $\Gamma$ (0.0, 0.0, 0.0) symmetry line in the Brillouin zone. The band gap with mBJ are 2.42 eV ($Rb_2SnBr_6$) and 0.84 eV ($Rb_2SnI_6$) which are somewhat larger than that calculated with PBE-GGA. When we include the spin-orbit coupling (SOC), the computed band gap for $Rb_2SnI_6$ reduces by 0.16 eV while for Br-containing compound, no significant changes have been observed. For $Rb_2SnI_6$, a density functional theory based HSE06+SOC calculations shows a direct band gap of 1.13 eV and 1.32 eV for tetragonal and monoclinic phases respectively[68]. However, UV-visible analysis shows an optical band gap of 1.32 eV, which is to some degree larger than our calculated band gap[68]. For these compounds, the valence band is mainly derived from Sn-5$s$ and X-5$p$ anti-bonding orbitals whereas the conduction band is derived entirely from Sn-5$p$ anti-bonding orbitals as shown in

Figure 8(a, b). For the Pd-based compounds, $Rb_2PdCl_6$, $Rb_2PdBr_6$, and $Rb_2PdI_6$, the computed band structures are shown in Figure 6. The plots reveal indirect band gap character for all the three compounds between the VBM and CBM at $\Gamma$ (0.0, 0.0, 0.0) and $X$ (0.0, 0.5, 0.5) symmetry lines. We see from the figure that with fixed A and B site cations, the band gap decreases by replacing Cl with Br and I. This trend may be due to the decrease of electronegativity difference between B-site elements (Pd@2.2) and halide ions (Cl@3.16, Br@2.96, and I@2.66) which increases the Pd-X covalent strength and make the valence band more dispersive (shown in Figure 6). This, in turn, pushes the Pd-X anti-bonding orbitals which form VBM higher in energy and results smaller band gap. Here the effect of SOC is not observed for $Rb_2PdCl_6$ and $Rb_2PdBr_6$, however, for $Rb_2PdI_6$ perovskite, the band gap fall shorter by 0.08 eV. The PBE-GGA method predicts the $Rb_2PdI_6$ compound as metal, however, both the mBJ and mBJ+SOC predicts it a narrow band gap semiconductor. The chosen k-path for this series of the compounds is $\Gamma$-$X$-$W$-$K$-$\Gamma$-$L$-$U$-$W$-$L$-$K$-$U$-$X$. For such k-path the valence band near the Fermi level have triply degenerated states. The bands are nearly parabolic at the $\Gamma$-point which is helpful to enhance the carrier mobility. The calculated density of states (DOS) shown in Figure 8(c-e) depicts that the valence band is primarily formed due to the mix contribution of Pd-$d$ and Cl/Br/I-$p$ orbitals and the conduction band is formed due to the Cl/Br/I-$p$ orbitals. Some bonding interaction can be seen in the lower part of the VBM between Pd-$d$ and Cl/Br/I-$p$ states. Strikingly, in Pd-based compounds there are triply degenerated states at VBM along the $\Gamma$-$X$ directions. Such states around VBM have a profound effect on the thermoelectric properties.

Finally, we consider $Cs_2PtCl_6$, $Cs_2PtBr_6$, and $Cs_2PtI_6$ compounds. The band structure plots for this series of compounds are displayed in Figure 7. One can see from the plots that for the same Cs cation, the band gap is found to increase gradually in the order such that the band gap for $Cs_2PtCl_6$ > $Cs_2PtBr_6$ > $Cs_2PtI_6$, consistent with the trend observed in $MAPbX_3$ (X = Cl, Br, I)[69]. Among these three compounds, $Cs_2PtCl_6$ perovskite possesses a direct band gap character (Figure 7) at the X symmetry line of the Brillouin Zone. The calculated direct gap ($E_g^{X-X}$) is 2.91 eV (mBJ) and 2.71 eV (mBJ+SOC), respectively. Inspection of the electronic band structure of $Cs_2PtBr_6$ perovskite reveals an indirect band gap character with a band gap value of 2.23 eV (mBJ functional). The computed band gap ($E_g$) of $Cs_2PtBr_6$ including SOC is 2.21 eV. Here the effect of SOC on the computed band gap is quite small and does not alter the band gap value significantly. The calculated band gap fall short by 0.02 eV when SOC is considered. For $Cs_2PtI_6$ crystal, the computed band structure reveals an indirect energy band gap character at $\Gamma$ and $X$ symmetry lines. The band gap using mBJ and mBJ+SOC are 1.22 eV and 1.11 eV, respectively. In this series of compounds, the lower part of the valence band is predominantly formed by Cs-$p$ (~ –7 eV) state whereas VBM is mainly derived from Cl/Br/I-$p$ states as presented in Figure 8(f-h). The higher conduction bands are derived from Cs-$p$ and Pt-$d$ states respectively. Here $d$-orbital of Pt splits into twofold degenerate $e_g$ states and threefold degenerate $t_{2g}$ states. The Pt-$e_g$ states are located in the upper part of conduction band and hybridized with X-$p$ states whereas the Pt-$t_{2g}$ states are situated at the VBM. It is worth mentioning that the mBJ calculated band gap is in good agreement with an earlier reported value of $E_g^{\Gamma-X}$ =1.340 eV[8] determined with the hybrid HSE-SOC functional. Hence the use of mBJ is justified in the present work. Interestingly, we notice that among the eight semiconductors we investigated, Rb$_2$PdBr$_6$ (1.31 eV) and Cs$_2$PtI$_6$ (1.22 eV) have favorable band gap in the optimal range of 0.9–1.6 eV, suggesting that they are ideal candidates for use in single-junctions PSCs.

**Table 3.** Band gap calculated by different GGA XC-functionals for A$_2$BX$_6$ (A= Rb, Cs; B= Sn, Pd, Pt; X= Cl, Br, I) compounds.

| Compound | $E_{g(PBE)}$/eV | $E_{g(EV-GGA)}$/eV | $E_{g(GLLB-SC)}$/eV | $E_{g(mBJ)}$/eV | $E_{g(mBJ-SOC)}$/eV | Other works | Exp: |
|---|---|---|---|---|---|---|---|
| Rb$_2$SnBr$_6$ | 1.30 | 1.609 | 1.878 | 2.42($\Gamma-\Gamma$) | 2.42($\Gamma-\Gamma$) | | |
| Rb$_2$SnI$_6$ | 0.12 | 0.424 | 0.462 | 0.84($\Gamma-\Gamma$) | 0.68($\Gamma-\Gamma$) | 1.02[8] | 1.32[68] |
| Rb$_2$PdCl$_6$ | 1.28 | 1.454 | 1.751 | 2.20($\Gamma-X$) | 2.20($\Gamma-X$) | | |
| Rb$_2$PdBr$_6$ | 0.61 | 0.811 | 0.959 | 1.31($\Gamma-X$) | 1.31($\Gamma-X$) | | |
| Rb$_2$PdI$_6$ | 0.0 | 0.224 | 0.204 | 0.47($\Gamma-X$) | 0.39($\Gamma-X$) | | |
| Cs$_2$PtCl$_6$ | 2.01 | 2.137 | 2.538 | 2.91(X$-$X) | 2.71(X$-$X) | | |
| Cs$_2$PtBr$_6$ | 1.42 | 1.612 | 1.834 | 2.23($\Gamma-X$) | 2.21(X$-$X) | | |
| Cs$_2$PtI$_6$ | 0.651 | 0.65 | 0.906 | 1.22($\Gamma-X$) | 1.11($\Gamma-X$) | 1.340[8] | |

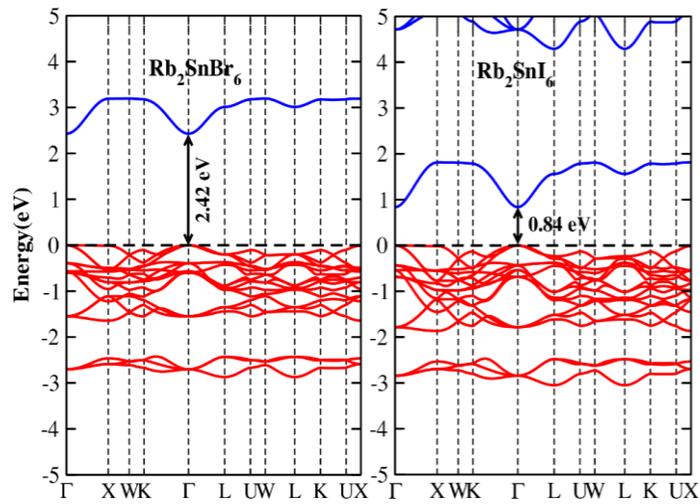

**Figure 5.** The calculated band structure of $Rb_2SnBr_6$ and $Rb_2SnI_6$ with TB-mBJ potential.

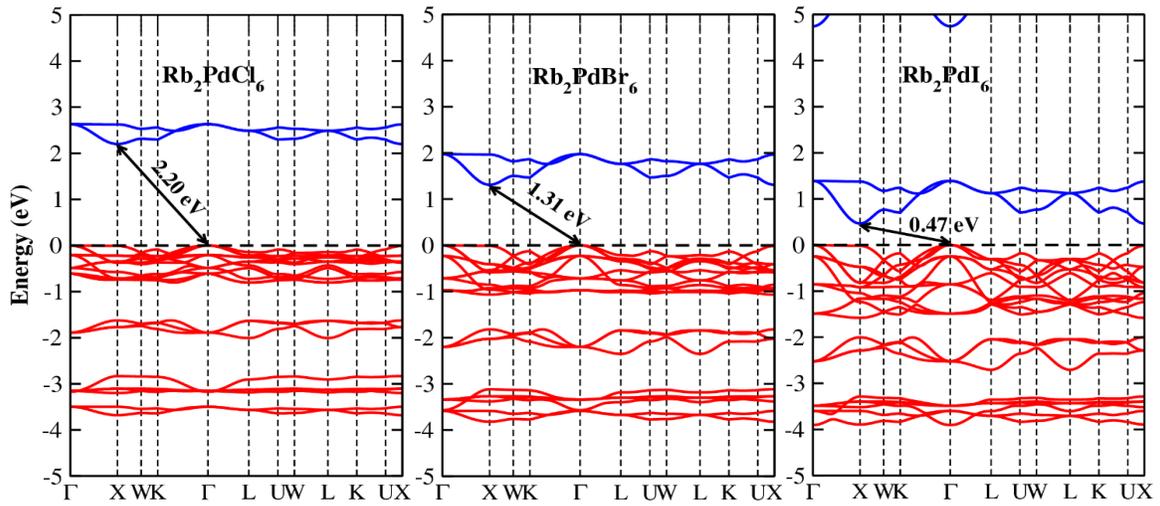

**Figure 6.** The calculated band structure of $Rb_2PdCl_6$, $Rb_2PdBr_6$, and $Rb_2PdI_6$ with TB-mBJ potential.

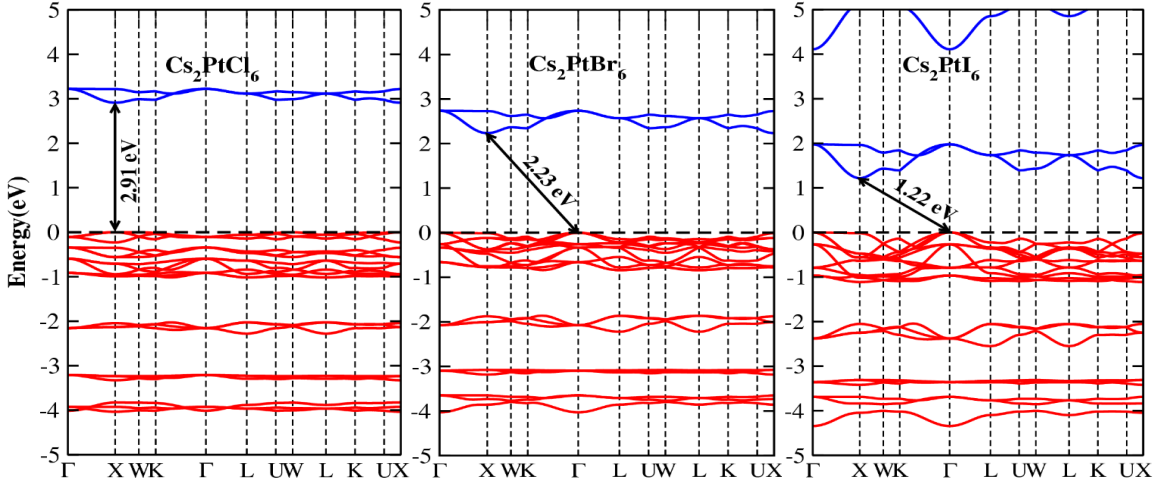

**Figure 7.** The calculated band structure of $Cs_2PtCl_6$, $Cs_2PtBr_6$, and $Cs_2PtI_6$ with TB-mBJ potential.

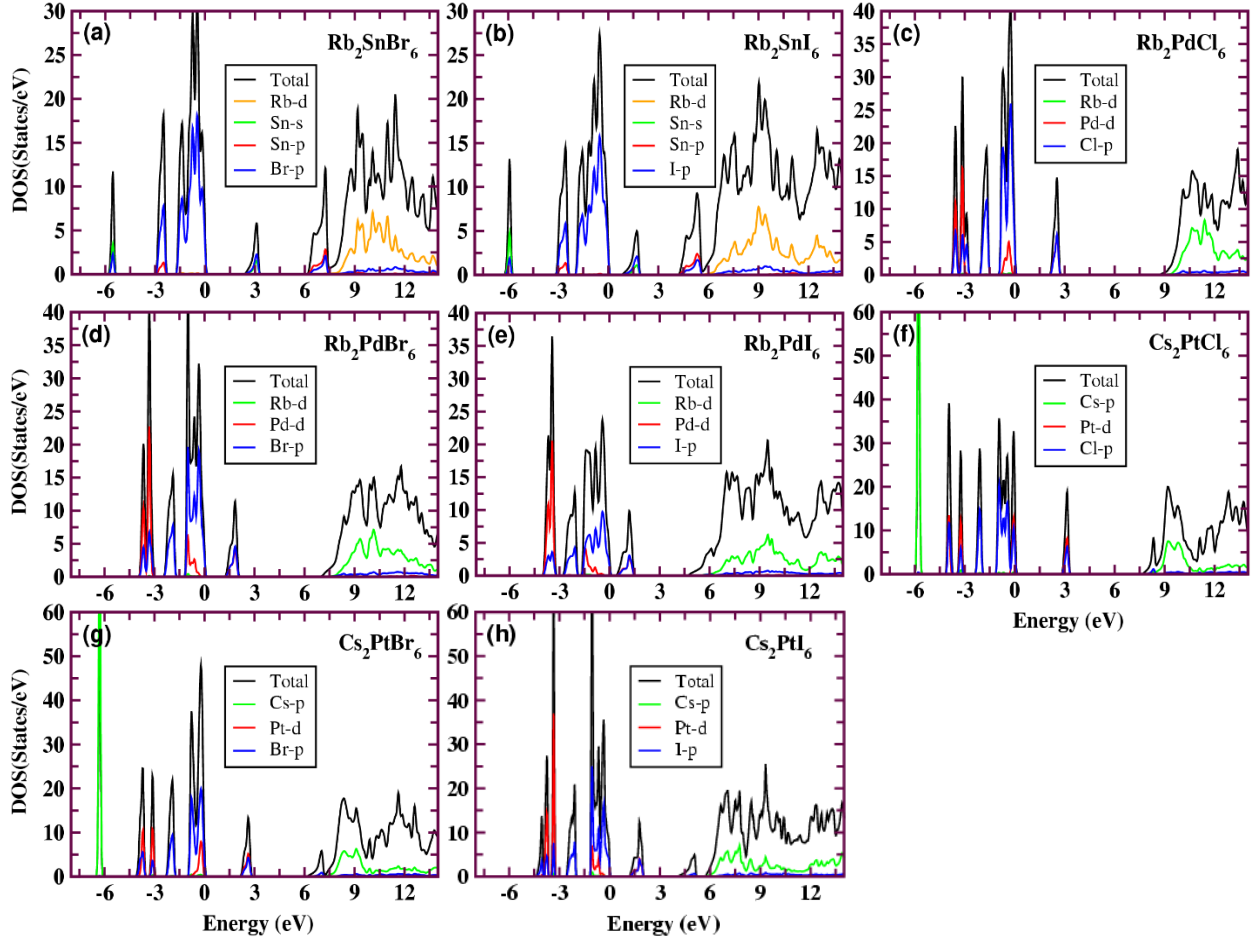

**Figure 8(a-h).** The total and partial density of states (DOS) for (a) $Rb_2SnBr_6$, (b) $Rb_2SnI_6$, (c) $Rb_2PdCl_6$, (d) $Rb_2PdBr_6$, (e) $Rb_2PdI_6$, (f) $Cs_2PtCl_6$, (g) $Cs_2PtBr_6$, and (h) $Cs_2PtI_6$, computed using mBJ potential. The Fermi level is set to 0 eV.

### 3.3. Optical Properties

To examine the performance of the $A_2BX_6$ (A= Rb and Cs; B = Sn, Pd, and Pt; and X= Cl, Br, and I) perovskite when exposed to light, we have performed the first principles calculations for the dielectric and optical properties. The properties like complex dielectric function, absorption spectra, optical conductivity, and reflectivity are investigated up to the photon energy of 10 eV to reveal the use of these materials in optoelectronic energy devices.

The optical response of a medium at all photon energies can be characterized by the dielectric function $\varepsilon(\omega)$. The real part of the dielectric function $\varepsilon_1(\omega)$ describe the energy stored in a medium while the imaginary part $\varepsilon_2(\omega)$ explains the absorption behavior of a crystal. Figures 9-11 depict various optical parameters for $A_2BX_6$ over the large energy range of 0–10 eV. The zero frequency limit, $\varepsilon_1(0)$, shown in Figure 9 give us the static dielectric constant of the compounds. For all these compounds, values of the static dielectric constant are listed in Table 4. The static dielectric constants are relatively small, which is an indication of the dielectric nature of the understudy compounds. Beyond the static point, $\varepsilon_1(\omega)$ increases and reaches a maximum value at certain photon energy for each of the compounds. The maximum peak of magnitude 9.9 (a.u) is observed for $Rb_2PdI_6$ at 1.15 eV. Similarly for the rest of the compounds, the maximum peak values are given in Table 4. After the maximum, a linear decrease in $\varepsilon_1(\omega)$ is observed for all these compounds, goes below zero at certain energy ranges and again it rises to the positive region. In negative energy range, the material lost their dielectric nature and exhibit metallic characteristics. Such metallic nature is due to the transition of electron from Cs-*p* and Sn-*p*, Pd-*p*, Pt-*p* states from valence band to the Cs-*d* and Sn/Pd/Pt-*p* states in the conduction band respectively. The frequency dependent imaginary part of the dielectric function $\varepsilon_2(\omega)$ explains the energy band structure and are presented in Figure 9 (bottom panel). The threshold for $Rb_2SnBr_6$ and $Rb_2SnI_6$ occur at 0.86 eV and 2.41 eV respectively. At this threshold energy, the optical transitions between VBM and CBM are direct as clear from the band structure plots (Figure 5). For $Rb_2PdCl_6$, $Rb_2PdBr_6$, and $Rb_2PdI_6$ the threshold energy are 2.12 eV, 1.33 eV, and 0.49 eV, respectively corresponding to the indirect optical transition between VBM and CBM. Similarly the thresholds for $Cs_2PtCl_6$ occur at 2.93 eV which reduces to 2.08 eV for $Cs_2PtBr_6$ and to 1.11 eV for $Cs_2PtI_6$. The optical transition are indirect for this series of the compounds as evident from the band structures. The observed thresholds are found in close agreement with our calculated band gap (Table 3) and are attributed to the inter-band electronic transitions. Above the threshold points, there are different peaks with some noticeable variations. It can be readily seen from the plots that the overall spectra of $\varepsilon_2(\omega)$ for $Rb_2BX_6$ (B = Sn and Pd; X = Cl, Br, and I) and $Cs_2PtX_6$ (X = Cl, Br, and I) is nearly identical, however, for $Rb_2PdI_6$ as well as for $Cs_2PtI_6$, the absorption peaks are more prominent as compare to the other compounds. Such peaks are due the electronic transitions from valance band (VB) states to the conduction band (CB) states. The maximum peak value of $\varepsilon_2(\omega)$ for all the compounds are recorded in Table 4.

The optical conductivities $\sigma(\omega)$ are also calculated and are presented in Figure 10. The optical conductance of $Rb_2SnX_6$ (X= Br, I) starts around 0.77 eV and 2.34 eV, for $Rb_2PdX_6$ (X= Cl, Br, I) compounds, the optical conduction starts around 0.56, 1.13, and 1.97 eV. Similarly, for cesium-based compounds $Cs_2PtX_6$ (X= Cl, Br, I), the optical conductivity 'σ' starts responding to the applied electromagnetic field around 2.7 eV, 2.3 eV, and 1.5 eV, respectively. Beyond these points the optical conductivity gradually increases by increasing the photon energy reach to maxima for each compound and then again decreases at various photon energies. As can be seen from the plots, the optical spectra is more prominent for iodide-based compounds and the maximum optical conductivity obtained for $Rb_2PdI_6$ and $Cs_2PtI_6$ is 7443 $\Omega^{-1}$ cm$^{-1}$ and 6985 $\Omega^{-1}$ cm$^{-1}$ respectively.

The absorption coefficient of any material is a measure of light penetration into a material at specific energy (wavelength)[70]. It also provides useful insights into the solar energy conversion efficiency for commercial application of materials in photovoltaics. The calculated absorption coefficients $\alpha(\omega)$ of the selected candidates are correspondingly shown in Figure 10 (bottom panel). The absorption of photon extends from around 2 eV to 10 eV. The absorption edges are located around 0.5−2.9 eV for different compounds which are in full agreement with the corresponding band gap as predicted by mBJ method. In absorption spectra there are several absorption peaks with increasing trend and are attributed to the electronic transitions from bonding states to the anti-bonding states, respectively. Among the eight compounds that were studied, the maximum absorption of 152.8 cm$^{-1}$ is found for $Rb_2PdI_6$ at about 9 eV. Therefore, Pd is found to be a better alternative of Pb in the inorganic perovskites solar cells. The replacement of the halogen ion by a lighter and smaller one i.e. I by Br and Br by Cl significantly alter the absorption spectra of the considered compounds. The absorption peaks shift towards the higher energy by changing the halogen ions in the same order. One can see from the plots that the absorption spectra is mostly concentrated in the ultraviolet energy range which makes these materials favorable for optical devices working in this range.

The reflectivity spectra 'R(ω)' for $A_2BX_6$ system have also been investigated and are collected in Figure 11. At zero vibration, the static reflectivity is found to be in the range of 5 % to 19 % for the selected compounds. However, the reflectivity increases with photon energy and become maximum around 5 eV. The static as well as the maximum reflectivity values for all the studied compounds are given in Table 4. The maximum reflectivity is found for $Rb_2PdCl_6$ and $Cs_2PtCl_6$ of magnitude 64% (4.7 eV) and 54% (5.6 eV) respectively. For $Rb_2PdCl_6$, $Rb_2PdI_6$, and $Cs_2PtI_6$, the reflectivity is in close agreement with that of $CH_3NH_3PbBr_3$ as reported by Park et al.[71]. The maximum reflectivity occurs at points where $\varepsilon_1(\omega)$ goes below zero. After the maximum, the reflectivity decreases with slight variation at various photon energies and then increases at high photon energies (6 eV). Such maximum reflectivity clearly suggest these compounds as favorable candidates for shielding and other protection purposes from high frequency radiations.

**Table 4.** Calculated static value of real part of dielectric function $\varepsilon_1(0)$, the maximum values of $\varepsilon_1(\omega)$, the maximum values of $\varepsilon_2(\omega)$, the static reflectivity R (0), the maximum peak values of R(ω) the maximum peak values of optical conductivity and absorption coefficient (i.e. $\sigma(\omega)_{max}$ $\alpha(\omega)_{max}$) for $A_2BX_6$ (A= Rb, Cs; B= Sn, Pd, Pt; X= Cl, Br, I) compounds.

|  | $Rb_2SnBr_6$ | $Rb_2SnI_6$ | $Rb_2PdCl_6$ | $Rb_2PdBr_6$ | $Rb_2PdI_6$ | $Cs_2PtCl_6$ | $Cs_2PtBr_6$ | $Cs_2PtI_6$ |
|---|---|---|---|---|---|---|---|---|
| $\varepsilon_1(0)$ | 2.68 | 3.88 | 3.10 | 4.32 | 6.76 | 2.53 | 3.27 | 4.61 |
| $\varepsilon_1(\omega)_{max}$ | 4.97 | 5.98 | 6.69 | 7.95 | 9.90 | 5.7 | 5.8 | 7.4 |
| $\varepsilon_2(\omega)_{max}$ | 3.98 | 5.82 | 6.57 | 6.86 | 9.54 | 6.38 | 6.05 | 7.24 |
| R(0) | 0.05 | 0.10 | 0.07 | 0.12 | 0.19 | 0.05 | 0.08 | 0.133 |
| R(ω) | 0.23 | 0.28 | 0.64 | 0.51 | 0.35 | 0.54 | 0.48 | 0.32 |
| $\sigma(\omega)_{max}$ | 5717 | 5963 | 5575 | 6266 | 7443 | 5815 | 5769 | 6985 |
| $\alpha(\omega)_{max}$ | 123.3 | 150.2 | 90.03 | 121 | 152.8 | 94.6 | 132.4.8 | 149.8 |

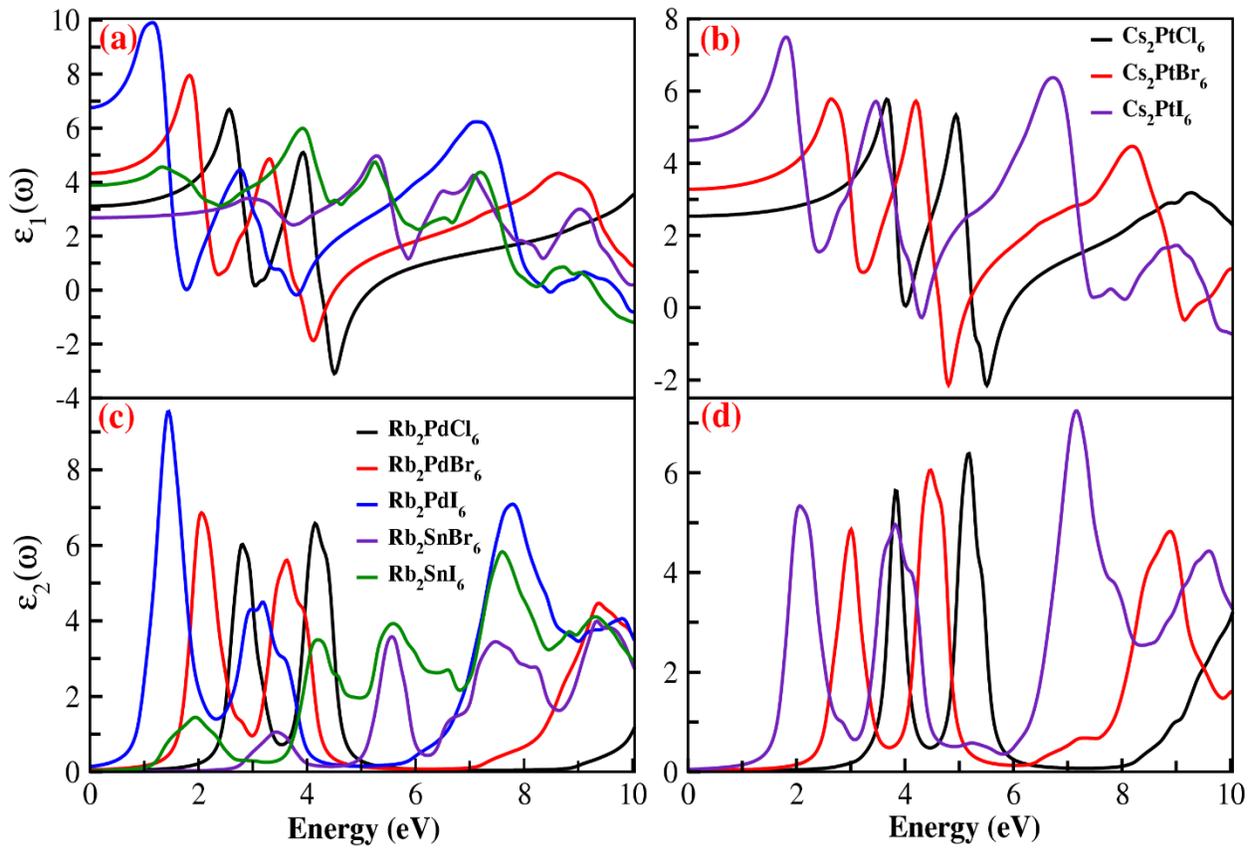

**Figure 9.** Real (top) and imaginary (bottom) parts of the dielectric function for $A_2BX_6$ (A= Rb, Cs; B= Sn, Pd, Pt; X= Cl, Br, I) computed using mBJ functional.

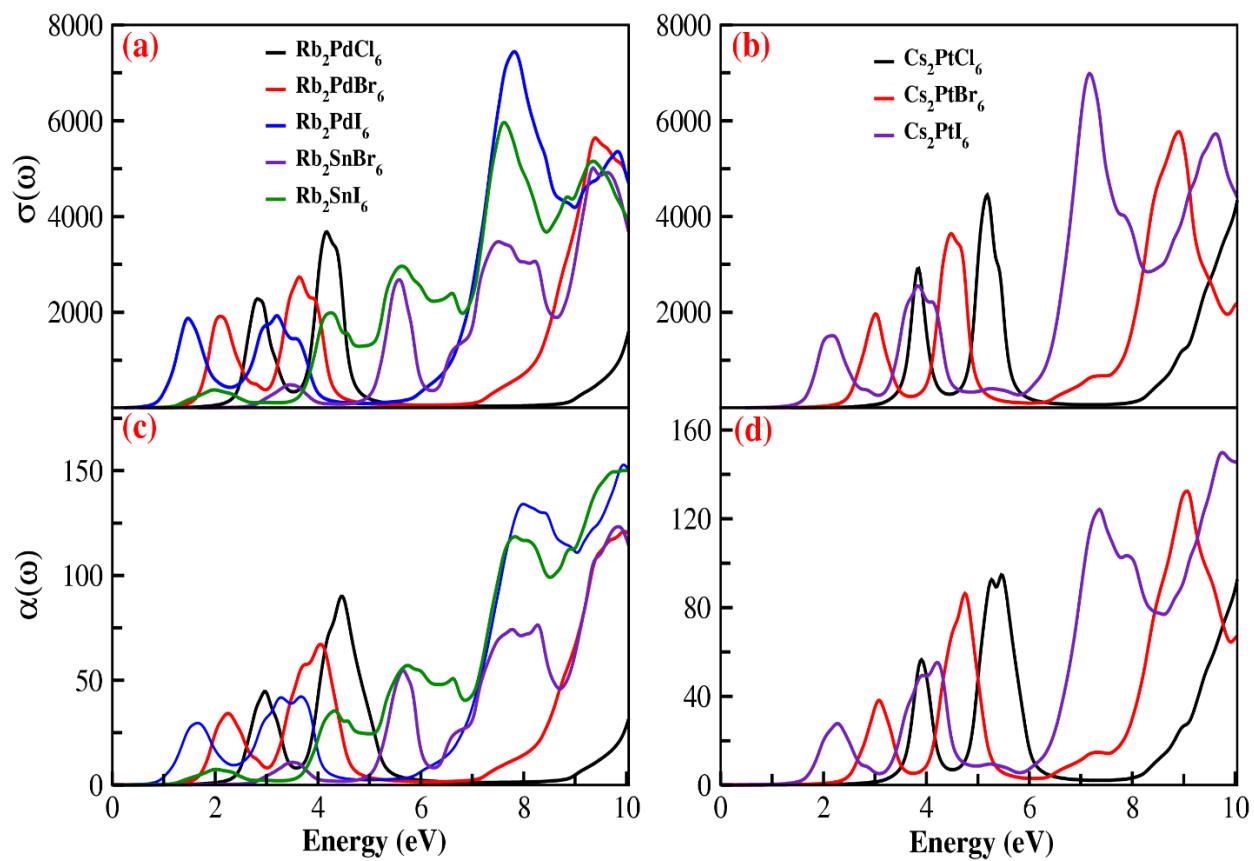

**Figure 10.** Optical conductivity (top) and absorption cofficient (bottom) for $A_2BX_6$ (A= Rb, Cs; B= Sn, Pd, Pt; X= Cl, Br, I) computed using mBJ functional.

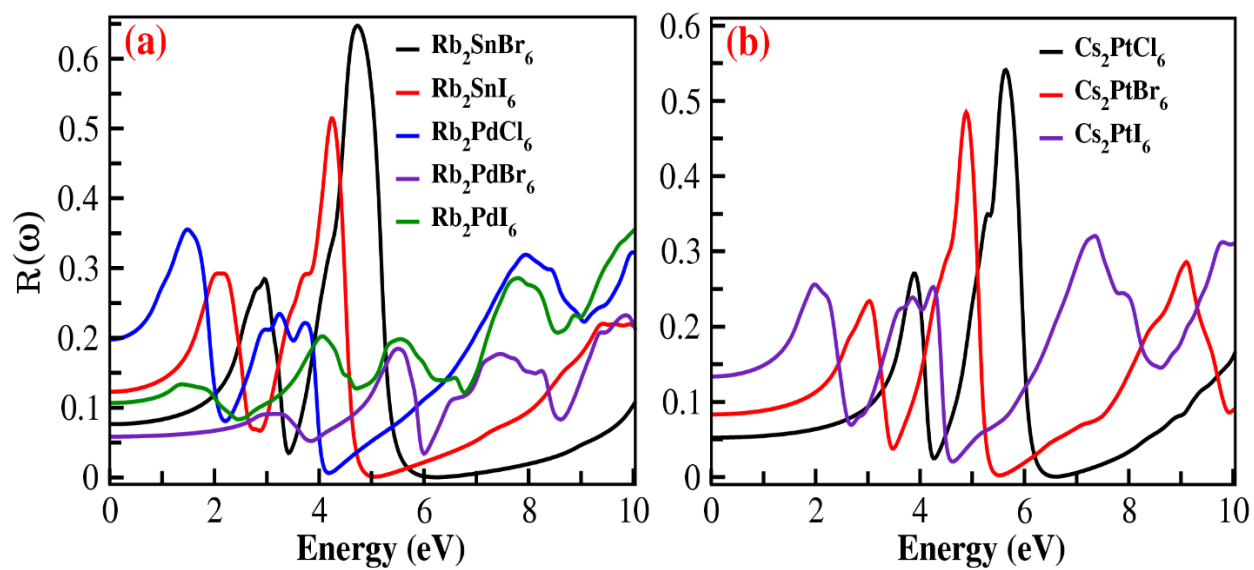

**Figure 11.** Reflectivity spectra *vs* incident photon energy of perovskites $A_2BX_6$ (A= Rb, Cs; B= Sn, Pd, Pt; X= Cl, Br, I).

## 4. CONCLUSIONS

In summary, we have performed first-principles calculations employing the mBJ potential to investigate the electronic structure as well as the optical properties of defect variant perovskites $A_2BX_6$ (A= Rb and Cs; B = Sn, Pd, and Pt; and X= Cl, Br, and I). The structural analysis shows a monotonic increase of the lattice constant and volume by changing the halide ion from Cl to Br and then to I which results in gradual increase in B-X bond length. The calculated enthalpies of formation for the investigated $A_2BX_6$ family are found to be negative, showing the thermodynamic stability of the selected candidates. The results show that all the considered compounds possess optimum electronic and optical properties as visible-light absorber materials for PV-applications. We also applied different exchange-correlation functionals, namely PBE-GGA, EV-GGA, GLLB+SC, GLLB+SC-SOC, and mBJ+SOC, which were essential in changing the band gap across the $A_2BX_6$ family. The calculated band gap varies most likely with the electronegativity difference of B- and X-site atoms. It is observed that VBM is triply degenerated and is mainly comprised of Cl/Br/I-$p$ states. Among the entire group of the compounds that were studied, ideal band gap were obtained for $Rb_2PdBr_6$ (1.31 eV) and $Cs_2PtI_6$ (1.22 eV) in the optimal range of 0.9−1.6 eV. This indicates that both the compounds can become the potential candidates for single-junction solar cells in the future. We also calculated different optical properties, most importantly, the complex dielectric function and absorption coefficient which reveal the use of these materials in various optoelectronic applications. The compounds, $Rb_2PdBr_6$, $Rb_2PdI_6$, and $Cs_2PtI_6$, possess suitable band gap and relatively high optical absorption as compared to the other members of this series of the compounds. Overall, our results provide key directions to promote the use of low cost lead free defect variant-perovskites in highly efficient photovoltaic and optoelectronic applications.

## ACKNOWLEDGMENTS

This work is supported by DST-SERB for the SERB-National Postdoctoral Fellowship (Award No. PDF/ 2017/002876). The author Muhammad Faizan thank the high performance computing Center of Development of Advanced Computing, Pune, India and Jilin University, China.

**REFERENCES**

1. Kojima, A.; Teshima, K.; Shirai, Y.; Miyasaka, T., Organometal halide perovskites as visible-light sensitizers for photovoltaic cells. *Journal of the American Chemical Society* 2009, *131* (17), 6050-6051.
2. Lee, M. M.; Teuscher, J.; Miyasaka, T.; Murakami, T. N.; Snaith, H. J., Efficient hybrid solar cells based on meso-superstructured organometal halide perovskites. *Science* 2012, 1228604.
3. Kim, H.-S.; Lee, C.-R.; Im, J.-H.; Lee, K.-B.; Moehl, T.; Marchioro, A.; Moon, S.-J.; Humphry-Baker, R.; Yum, J.-H.; Moser, J. E., Lead iodide perovskite sensitized all-solid-state submicron thin film mesoscopic solar cell with efficiency exceeding 9%. *Scientific reports* 2012, *2*, 591.
4. NREL, 2018.
5. Giorgi, G.; Fujisawa, J.-I.; Segawa, H.; Yamashita, K., Small photocarrier effective masses featuring ambipolar transport in methylammonium lead iodide perovskite: a density functional analysis. *The journal of physical chemistry letters* 2013, *4* (24), 4213-4216.
6. Frohna, K.; Deshpande, T.; Harter, J.; Peng, W.; Barker, B. A.; Neaton, J. B.; Louie, S. G.; Bakr, O. M.; Hsieh, D.; Bernardi, M., Inversion symmetry and bulk Rashba effect in methylammonium lead iodide perovskite single crystals. *Nature communications* 2018, *9* (1), 1829.
7. Stranks, S. D.; Eperon, G. E.; Grancini, G.; Menelaou, C.; Alcocer, M. J.; Leijtens, T.; Herz, L. M.; Petrozza, A.; Snaith, H. J., Electron-hole diffusion lengths exceeding 1 micrometer in an organometal trihalide perovskite absorber. *Science* 2013, *342* (6156), 341-344.
8. Cai, Y.; Xie, W.; Ding, H.; Chen, Y.; Thirumal, K.; Wong, L. H.; Mathews, N.; Mhaisalkar, S. G.; Sherburne, M.; Asta, M., Computational Study of Halide Perovskite-Derived A2BX6 Inorganic Compounds: Chemical Trends in Electronic Structure and Structural Stability. *Chemistry of Materials* 2017, *29* (18), 7740-7749.
9. Song, Z.; Abate, A.; Watthage, S. C.; Liyanage, G. K.; Phillips, A. B.; Steiner, U.; Graetzel, M.; Heben, M. J., Perovskite Solar Cell Stability in Humid Air: Partially Reversible Phase Transitions in the PbI2‐CH3NH3I‐H2O System. *Advanced Energy Materials* 2016, *6* (19), 1600846.
10. Nagabhushana, G.; Shivaramaiah, R.; Navrotsky, A., Direct calorimetric verification of thermodynamic instability of lead halide hybrid perovskites. *Proceedings of the National Academy of Sciences* 2016, *113* (28), 7717-7721.
11. Hailegnaw, B.; Kirmayer, S.; Edri, E.; Hodes, G.; Cahen, D., Rain on methylammonium lead iodide based perovskites: possible environmental effects of perovskite solar cells. *The journal of physical chemistry letters* 2015, *6* (9), 1543-1547.
12. Bass, K. K.; McAnally, R. E.; Zhou, S.; Djurovich, P. I.; Thompson, M. E.; Melot, B. C., Influence of moisture on the preparation, crystal structure, and photophysical properties of organohalide perovskites. *Chemical Communications* 2014, *50* (99), 15819-15822.
13. Mosconi, E.; Azpiroz, J. M.; De Angelis, F., Ab initio molecular dynamics simulations of methylammonium lead iodide perovskite degradation by water. *Chemistry of Materials* 2015, *27* (13), 4885-4892.

**TOC Graphics**

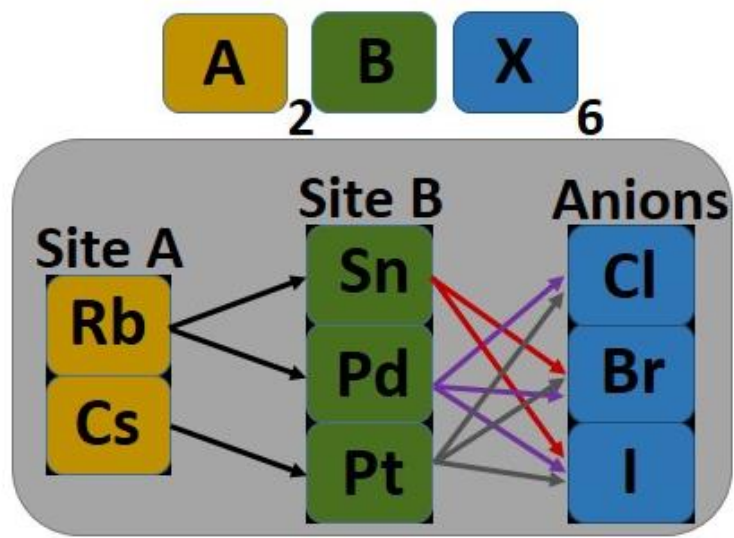